\documentclass[]{spie}

\usepackage{amsmath,amsfonts,amssymb}
\usepackage{graphicx}
\usepackage[colorlinks=true, allcolors=blue]{hyperref}

\title{FINER: development of the wideband millimeter-wave receiver system and preparations for first light on the Large Millimeter Telescope}

\author[a]{Akio Taniguchi}
\author[b]{Yoichi Tamura}
\author[c]{Takeshi Sakai}
\author[b]{Masato Hagimoto}
\author[d]{Takuya Hashimoto}
\author[e]{Akio K. Inoue}
\author[f,g]{Shun Ishii}
\author[f]{Haoran Kang}
\author[b]{Masato Kato}
\author[f,g]{Ryohei Kawabe}
\author[h]{Kotaro Kohno}
\author[f]{Takafumi Kojima}
\author[b]{Kianhong Lee}
\author[f]{Sho Masui}
\author[b]{Akinobu Miyake}
\author[i]{Taku Nakajima}
\author[f,g]{Tai Oshima}
\author[f]{Wenlei Shan}
\author[a]{Tatsuya Takekoshi}
\author[j]{Kunihiko Tanaka}
\author[f,g]{Kotomi Taniguchi}
\author[a]{Kaoru Tsuji}
\author[k]{Issei Watanabe}
\author[l]{Tom J. L. C. Bakx}
\author[m]{Yoshinobu Fudamoto}
\author[n]{Kazuyuki Fujita}
\author[o]{Yuichi Harikane}
\author[f,g]{Bunyo Hatsukade}
\author[p]{David H. Hughes}
\author[q]{Takahiro Iino}
\author[n]{Yuki Kimura}
\author[r]{Hiroyuki Maezawa}
\author[f,g]{Yuichi Matsuda}
\author[e]{Ken Mawatari}
\author[s]{Shunichi Nakatsubo}
\author[t]{Hideo Sagawa}
\author[u]{Peter F. Schloerb}
\author[v]{Shigeru Takahashi}
\author[h]{Akiyoshi Tsujita}
\author[b]{Hideki Umehata}
\author[u]{Min S. Yun}

\affil[a]{Kitami Institute of Technology, 165 Koen-cho, Kitami, Hokkaido 090-8507, Japan}
\affil[b]{Nagoya University, Furocho, Chikusa-ku, Nagoya, Aichi 464-8602, Japan}
\affil[c]{The University of Electro-Communications, 1-5-1 Chofugaoka, Chofu, Tokyo 182-8585, Japan}
\affil[d]{University of Tsukuba, Tsukuba, Ibaraki 305-8571, Japan}
\affil[e]{Waseda University, 3-4-1 Okubo, Shinjuku-ku, Tokyo 169-8555, Japan}
\affil[f]{National Astronomical Observatory of Japan, 2-21-1 Osawa, Mitaka, Tokyo 181-8588, Japan}
\affil[g]{The Graduate University for Advanced Studies (SOKENDAI), 2-21-1 Osawa, Mitaka, Tokyo 181-8588, Japan}
\affil[h]{The University of Tokyo, 2-21-1 Osawa, Mitaka, Tokyo 181-0015, Japan}
\affil[i]{Suwa University of Science, 5000-1 Toyohira, Chino, Nagano 391-0292, Japan}
\affil[j]{Keio University, 3-14-1 Hiyoshi, Yokohama, Kanagawa 223-8522, Japan}
\affil[k]{National Institute of Information and Communications Technology, 4-2-1 Nukui-Kitamachi, Koganei, Tokyo 184-8795, Japan}
\affil[l]{Chalmers University of Technology, Gothenburg SE-412 96, Sweden}
\affil[m]{Chiba University, 1-33 Yayoicho, Inage-ku, Chiba, Chiba 263-8522, Japan}
\affil[n]{Hokkaido University, Kita-19, Nishi-8, Kita-ku, Sapporo, Hokkaido 060-0819, Japan}
\affil[o]{The University of Tokyo, 5-1-5 Kashiwanoha, Kashiwa, Chiba 277-8582, Japan}
\affil[p]{Instituto Nacional de Astrofísica, Óptica y Electrónica, Luis Enrique Erro 1, Tonantzintla C.P. 72840, Puebla, México}
\affil[q]{University of Yamanashi, 4-4-37 Takeda, Kofu, Yamanashi 400-8510, Japan}
\affil[r]{Osaka Metropolitan University, 1-1 Gakuen-cho, Naka-ku, Sakai, Osaka 599-8531, Japan}
\affil[s]{Japan Aerospace Exploration Agency, 3-1-1 Yoshinodai, Chuo-ku, Sagamihara, Kanagawa 252-5210, Japan}
\affil[t]{Kyoto Sangyo University, Motoyama, Kamigamo, Kita-ku, Kyoto 603-8555, Japan}
\affil[u]{University of Massachusetts, Amherst, MA 01003, USA}
\affil[v]{University of Hyogo, 407-2 Nishigaichi, Sayo-cho, Sayo-gun, Hyogo 679-5313, Japan}

\authorinfo{
    Further author information: (Send correspondence to Akio Taniguchi)\\
    Akio Taniguchi: E-mail: a-taniguchi@mail.kitami-it.ac.jp\\
    Yoichi Tamura: E-mail: ytamura@nagoya-u.jp\\
    Takeshi Sakai: E-mail: takeshi.sakai@uec.ac.jp
}

\begin{document}
\maketitle

\begin{abstract}
The recent discovery of an excess of luminous galaxies in the early Universe necessitates sensitive and wideband millimeter spectroscopy to understand their rapid growth.
To address this, we present the development of the Far-Infrared Nebular Emission Receiver (FINER) for the Large Millimeter Telescope (LMT).
The FINER frontend comprises two receivers covering 120--350 GHz (corresponding to ALMA Bands 4+5 and 6+7).
The warm optics are designed to enable simultaneous two-band observations.
Combined with the 10.24-GHz-wide digital spectrometer array, the system aims to deliver an instantaneous bandwidth approximately five times wider than current ALMA capabilities.
We report that the 210--350 GHz receiver has already achieved commissioning-level performance, with sideband rejection further enhanced by the digital sideband separation technique.
With installation expected in 2026, we discuss parallel preparations, including integrated testing and commissioning plans for first-look targets.
\end{abstract}


\section{THE FAR-INFRARED NEBULAR EMISSION RECEIVER PROJECT}

The recent discovery by \textit{JWST} of an excess of massive galaxy candidates challenges our understanding of galaxy formation in the epoch of reionization and beyond\cite{Harikane+2023}.
Millimeter and submillimeter spectroscopic observations of redshifted far-infrared emission lines, particularly the \textsc{[O\,iii]} 88 micron and \textsc{[C\,ii]} 158 micron lines, are key to uncovering the physical mechanisms behind their rapid growth\cite{Inoue+2016,Hashimoto+2018,Tamura+2019,Bakx+2020,Hagimoto+2025,Nakazato+2026}.
To this end, the Far-Infrared Nebular Emission Receiver (FINER) project\cite{Tamura+2024} is realizing a wideband millimeter receiver system for the Large Millimeter Telescope (LMT)\cite{Hughes+2020}.
Combining the five times wider bandwidth than the current ALMA with the superb atmospheric condition and 40\% of ALMA’s light collecting area offered by the LMT, it is expected to deliver the most sensitive millimeter spectral scanning capability in the northern hemisphere and offer unique synergies with the upcoming large all-sky samples from \textit{JWST}, \textit{Euclid}, and \textit{Roman}.

FINER consists of two wideband, dual-polarization, sideband-separating (2SB) superconductor-insulator-superconductor (SIS) receivers covering 120--210 GHz and 210--350 GHz frequency ranges, which correspond to ALMA Bands 4+5 and 6+7, respectively\footnote{Throughout this proceeding, we refer to the FINER receivers using the band definitions of ALMA.}.
Applying high critical current density SIS mixer fabrication techniques that can also be utilized in the ALMA Wideband Sensitivity Upgrade\cite{Carpenter+2023}, FINER offers an intermediate frequency (IF) bandwidth of 3--21 GHz for both upper and lower sidebands in each polarization\cite{Kang+2024}.
The IF bandwidths are fully and simultaneously covered by the 10.24 GHz-wide digital spectrometer array (DRS4; Elecs Industry Co., Ltd.) equipped with the digital sideband separation (DSBS) functionality to approach uniform sensitivity even within such wide bandwidths where the atmospheric transmission cannot be considered constant\cite{Hagimoto+2024}.

\section{INTEGRATED SYSTEM TESTING AND COMMISSIONING PLANS}

Toward the installation, commissioning and science verification of FINER on the LMT expected in 2026, we are advancing parallel preparations; integrated system testing in the laboratory, optimization of observation strategies utilizing a data-scientific noise removal method, and selection of the first-look targets at $z > 10$.

\begin{figure}[h]
    \centering
    \includegraphics[width=\textwidth]{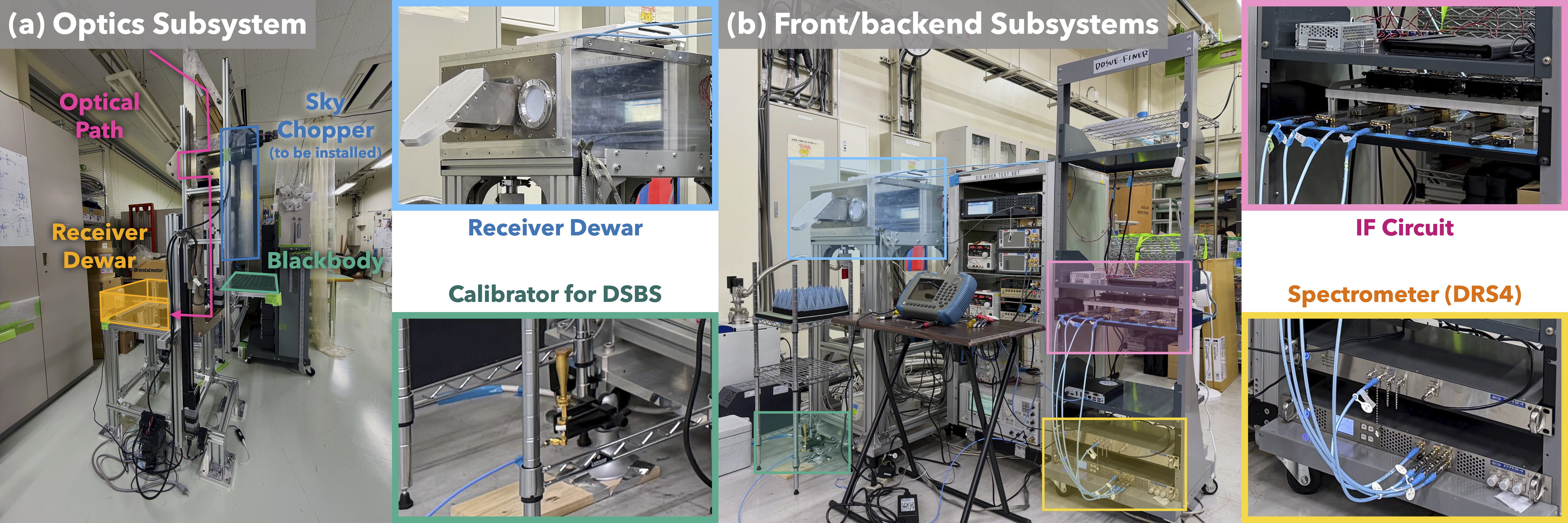}
    \caption{
        Photos of the current laboratory setup for the integrated system testing of FINER at NAOJ/ATC.
        (a) The optics subsystem, where the orange box indicates the location of the receiver dewar to be installed.
        It will be replaced with the existing Band 4 receiver (B4R)\cite{Kawabe+2025,Yonetsu+2025} at the receiver room of the LMT.
        (b) The frontend and backend subsystems, where the Band 6+7 receiver is currently installed in the dewar.
        The DC--10.24 and 10.24--20.48 GHz IF signals of the both sidebands are fed into the DRS4 spectrometer array.
        A continuous wave signal is injected into the receiver to estimate the complex gain coefficients to compensate for the amplitude and phase imbalances between the two sidebands (DSBS)\cite{Finger+2013,Rodriguez+2018}.
    }
    \label{fig:photos}
\end{figure}

We have been assembling the optics, frontend, and backend subsystems at the Advanced Technology Center (ATC) of the National Astronomical Observatory of Japan (NAOJ) and conducting integrated system testing.
Figure~\ref{fig:photos} shows the photo of the current laboratory setup. We have designed the warm optics to enable simultaneous two-band (two-beam; separation of 1\,arcmin) observations by switching the on- and off-source positions, which will be upgraded by the implementation of a sky chopper, a rotating mirror that optically switches the beam directions\cite{Moerman+2025}, for an efficient on-off switching without telescope movement.
Meanwhile, we have assembled the Band 6+7 receiver and confirmed commissioning-level performance.
We demonstrated the receiver noise temperature ($T_{\mathrm{RX}}$) of $\sim$100\,K and the analog sideband rejection ratios (SRR) of $\sim$10\,dB, which is further enhanced to $\sim$20\,dB utilizing the DSBS functionality of DRS4 (details to be presented by Kang et al., this conference, 14156-27).
For the Band 4+5 receiver, waveguide component fabrication is completed, while wideband SIS mixers are under fabrication.
During the commissioning and science verification, we will alternatively use a monolithic microwave integrated circuit (MMIC) receiver (125--163\,GHz)\cite{Shan+2026}.

\begin{figure}[h]
    \centering
    \includegraphics[width=\textwidth]{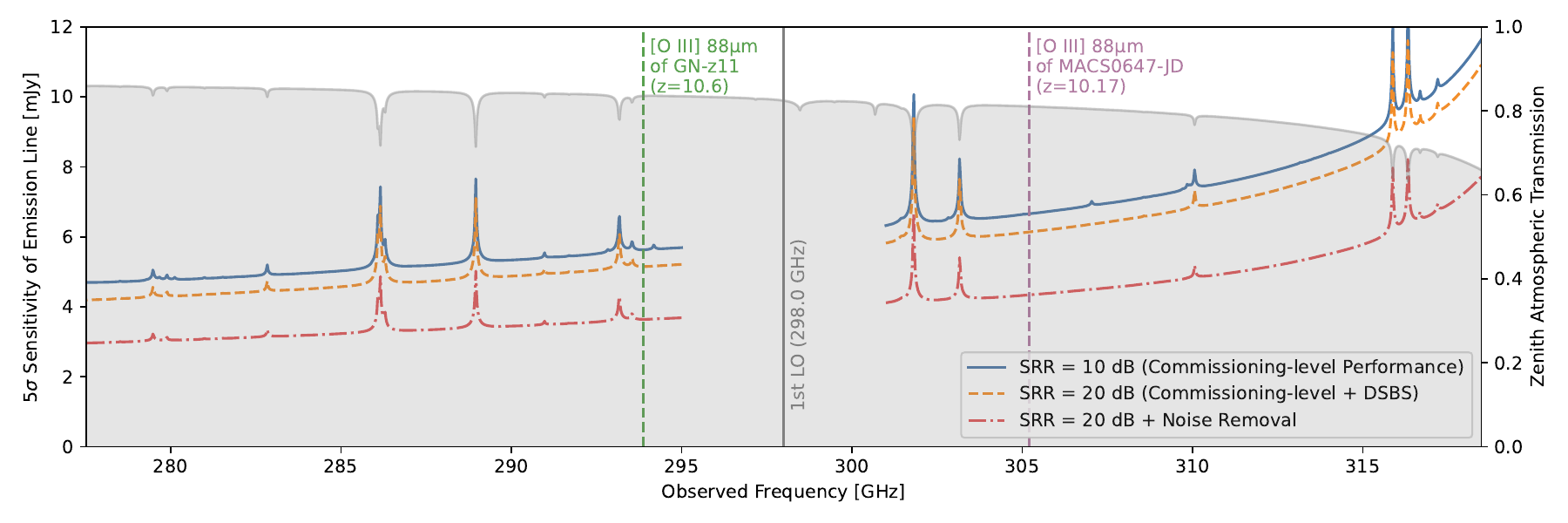}
    \caption{
        Estimated $5\sigma$ sensitivity of the FINER Band 6+7 receiver with an on-source integration time of 10 hours and a frequency setup targeting the \textsc{[O\,iii]} 88 micron lines at $z \approx 10$ (1st LO frequency = 298\,GHz).
        Throughout the calculation, we assume the commissioning-level performance of the receiver ($T_{\mathrm{RX}} = 100$\,K, SRR = 10\,dB) and the typical LMT site conditions (precipitable water vapor = 2.0\,mm (see also the zenith atmospheric transmission curve), observation elevation = 60\,deg, LMT's surface accuracy = 100\,$\mu$m\cite{Kawabe+2025}).
        In addition to the sensitivity under the commissioning-level performance (blue curve), we also show the sensitivity improved by the DSBS technique (SRR = 20\,dB; dashed orange curve) and that further improved by the data-scientific noise-removal method\cite{Taniguchi+2021} (by a factor of $\sim\sqrt{2}$; dash-dotted red curve).
    }
    \label{fig:sensitivity}
\end{figure}

During the commissioning and science verification, we plan to observe standard sources such as the planets, bright quasars, and Galactic star-forming regions for establishing the pointing model and estimating the aperture efficiency in the FINER bands, and to observe CO lines of bright dusty star-forming galaxies for the verification of line detection capability.
We then plan to conduct deep spectral observation targeting the \textsc{[O\,iii]} 88 micron and/or \textsc{[C\,ii]} 158 micron lines for a few first-look targets at $z > 10$.
Figure~\ref{fig:sensitivity} shows the expected $5\sigma$ sensitivity of the FINER Band 6+7 receiver with an on-source integration time of 10 hours and a frequency setup targeting the \textsc{[O\,iii]} 88 micron lines at $z \approx 10$, suggesting the detection of the emission lines at a several mJy such as GN-z11\cite{Bunker+2023}.
We plan to further optimize the spectral observations utilizing a data-scientific noise removal method\cite{Taniguchi+2021} demonstrated with the predecessor Band 4 receiver (B4R)\cite{Kawabe+2025,Yonetsu+2025}, which will improve the sensitivity by a factor of $\sim\sqrt{2}$.
Figure~\ref{fig:sensitivity} also shows the expected $5\sigma$ sensitivity with the noise removal method, enabling us to detect the emission lines with a shorter integration time or at a higher signal-to-noise ratio.

\acknowledgments
This work was financially supported by JSPS KAKENHI (Nos. 22H04939, 22J21948, 22KJ1598, 23K20035), NAOJ Joint Development Research Programs (Nos. 1901-0101, 2001-0104), and the Grant for Joint Research Program of the Institute of Low Temperature Science, Hokkaido University (23G037, 24G036, 25G034, 26G033).
We acknowledge the use of the facilities at the Advanced Technology Center (ATC) of the National Astronomical Observatory of Japan (NAOJ).
The LMT project is a joint effort between the Instituto Nacional de Astr\'{o}fisica, \'{O}ptica, y Electr\'{o}nica (INAOE) and the University of Massachusetts at Amherst (UMASS).

\bibliography{references}
\bibliographystyle{spiebib}
\end{document}